\documentclass[conference,a4paper]{IEEEtran}
\usepackage{float}

\usepackage{amssymb,amsmath}
\usepackage{cite}
\usepackage{graphicx,subfigure}
\usepackage{psfrag}
\usepackage{url}
\usepackage{bbm}
\usepackage{amsmath}
\usepackage{amsthm}
\usepackage{mathtools}
\usepackage[absolute,overlay]{textpos}
\usepackage{dsfont}
\usepackage{setspace}
\usepackage{float}
\usepackage{color}
\usepackage[utf8]{inputenc}
\usepackage[english]{babel}
\newtheorem{remark}{Remark}
\makeatletter
\renewcommand{\fnum@figure}{Fig. \thefigure}
\makeatother

\begin{document}

\title{On the SIR Meta Distribution in Massive MTC Networks with Scheduling and Data Aggregation}
\author{Nelson J. Mayedo Rodríguez, Onel L. Alcaraz López, Hirley Alves, and Matti Latva-aho\\
    \IEEEauthorblockA{
        Centre for Wireless Communications (CWC), University of Oulu, Oulu, Finland\\
    }
    E-mail: \{nelson.mayedorodriguez, onel.alcarazlopez, hirley.alves, matti.latva-aho\}@oulu.fi
}
\maketitle

\begin{abstract}
Data aggregation is an efficient approach to handle the congestion introduced by a massive number of machine type devices (MTDs). The aggregators not only collect data but also implement scheduling mechanisms to cope with scarce network resources.
  We use the concept of meta distribution (MD) of the signal-to-interference ratio (SIR) to gain a complete understanding of the per-link reliability and describe the performance of two scheduling methods for data aggregation of machine type communication (MTC): random resource scheduling (RRS) and channel-aware resource scheduling (CRS). The results show the fraction of users in the network that achieves a target reliability, which is an important aspect to consider when designing wireless systems with stringent service requirements.
\end{abstract}
%
\section{Introduction}
Nowadays, machine-type communication (MTC) applications\textemdash smart building and surveillance, smart cities, smart grid, remote maintenance and monitoring systems\textemdash have a significant impact on society and constitute an important factor in economic development. These applications fall under the Internet of Things (IoT) umbrella, which envisions connectivity for a huge number of heterogeneous, low-complexity, low-power and low-storage devices (e.g., sensors, actuators, appliances) communicating between each other through the Internet without human intervention. This variety of system requirements brings numerous challenges in connectivity and efficient communication management \cite{akpakwu2017survey}.

\begin{figure}[t!]
    \centering
    \includegraphics[width=0.45\textwidth]{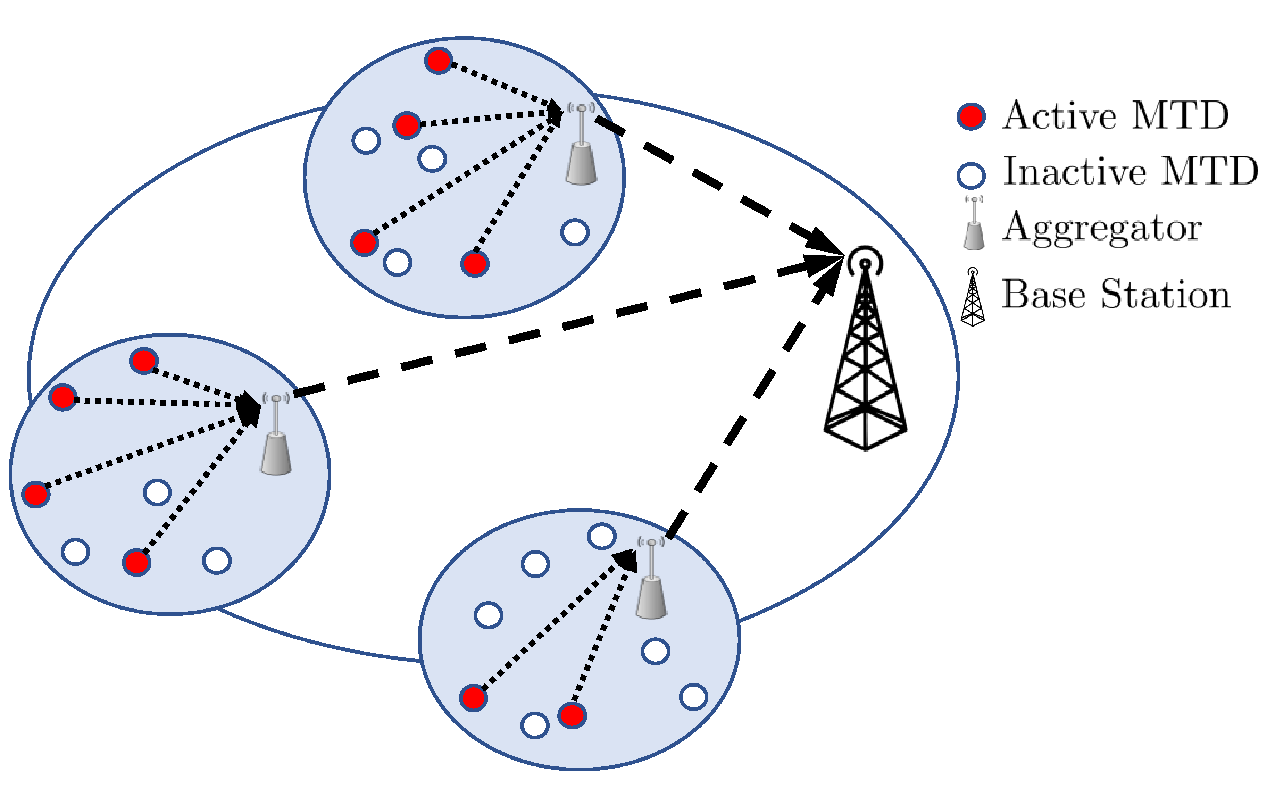}
    \caption{Typical MTC scenario with data aggregation.}
    \label{fig1}
    \vspace{-2mm}
\end{figure}

A promising way to enable a massive number of simultaneously connected devices relies on the concept of data aggregation, which means that the traffic coming from machine-type devices (MTDs) is first collected by nodes called aggregators, which then relay the data to the core network. As shown in Fig.\ref{fig1}, this structure \cite{7736615}: $\romannumeral 1)$ shortens the distance in the communication while diminishing the power consumption of the MTDs; $\romannumeral 2)$ reduces the number of connections to the core, thus decreasing the congestion; and $\romannumeral 3)$ extends network coverage. Several articles have recently investigated and exploited the advantages of data aggregation in massive MTC (mMTC). For example, the authors of \cite{7736615} presented experimental results that quantify the signaling load reduction as the number of aggregators grows in two practical scenarios: smart metering and vehicular sensing. An efficient network slicing data aggregation scheme is introduced in [3] for MTC applications in 5G networks. In contrast to conventional strategies that only consider the device location, this solution also assesses the latency requirement to effectively increase the network capacity and reduce the collision probability and average access delay of devices. In \cite{ullah2020efficient}, the authors proposed a data aggregation scheme based on the clustering of sensor nodes and extreme learning machine. This design employs a Kalman filter to transmit only accurate and condensed data to the base station. In \cite{hamzah2019energy}, a novel clustering technique based on fuzzy logic is suggested for cluster head selection and energy-efficient routing protocols in wireless sensor networks with data aggregation.

Because of the scarce network resources, aggregators can also perform different types of scheduling mechanisms. This allows incorporating some intelligence in the aggregator such that the network performance improves \cite{guo2017massive}, \cite{chang2012not}. For these reasons,
 some solutions have been proposed to optimize the data aggregation process. In \cite{5336978}, a scheduling mechanism is designed to minimize the latency of the aggregation phase. Meanwhile, two scheduling schemes: random resource scheduling (RRS) and channel-aware resource scheduling (CRS) are proposed in \cite{guo2017massive} and analyzed using stochastic geometry. The authors show that CRS and RRS schemes achieve similar performance as long as the available resources in the aggregator are not very limited, while CRS outperforms RRS when the number of MTDs requesting service exceeds the number of available channels. Later, these scheduling mechanisms were extended in \cite{lopez2018aggregation}, \cite{lopez2019hybrid} under non-orthogonal multiple access (NOMA) and imperfect successive interference cancellation.
 
 On the other hand, the meta distribution (MD) of the signal-to-interference ratio (SIR), which is the distribution of the success probability conditioned on the point process, has become an attractive tool to understand the behavior of large-scale cellular networks better. By characterizing the SIR MD, one can obtain useful insights on how the link success probabilities are distributed in a realization of the network. For example, the MD is used in \cite{ibrahim2019meta} to evaluate the reliability and latency of a cellular network with services coexisting in the  sub-6GHz and millimeter wave spectrum. Authors of \cite{kalamkar2019per} exploited the SIR MD to adjust the per-link reliability via rate control. The MD is further used in \cite{koufos2019meta} to describe the behavior of linear motorway vehicular ad hoc networks, where the vehicles' high speed makes the Poisson point process not suitable for system modeling. In \cite{salehi2018meta}, the MD allows the authors to optimize the power allocation in downlink NOMA with and without latency constraints.
 
 To the best of our knowledge, current literature lacks a framework that integrates the three aforementioned concepts: data aggregation, MD, and resource scheduling. Our objective is to fill this gap through the following contributions:
\begin{itemize}
\item We use the MD concept to fully characterize the uplink traffic performance in a Poisson network with data aggregation. We adopt the RRS and CRS scheduling schemes proposed in \cite{guo2017massive} to deal with the limited spectrum resources. However, we provide a more fine-grained performance characterization of a typical link. 
\item We present an accurate and simple closed-form expression for the SIR MD under the RRS scheme based on an approximation obtained from the relation between the geometric and the arithmetic mean, in contrast to the usually adopted moment inversion approach (Gil-Pelaez theorem) that implies the calculation of complex integrals.
\item Our results show that CRS can serve more devices than RRS for a common target reliability. Moreover, we conclude that the standard success probability does not guarantee quality-of-service (QoS) for any node in a network.
We also provide insights into the transmission rate required to keep a percentage of devices/users communicating with a target reliability for both scheduling schemes.
\end{itemize}
The rest of the paper is structured as follows.  Section \textrm{II} presents the system model and assumptions. Section \textrm{III} defines the SIR MD as a per-link performance metric. Sections \textrm{IV} and \textrm{V} study RRS and CRS scheduling schemes, respectively. Finally, Section \textrm{VI} presents numerical results and network design insights, while Section \textrm{VII} concludes the paper.
\section{System Model and Assumptions}
 We examine the uplink transmissions of a large-scale single-tire cellular network overlaid with spatially distributed aggregator nodes. One specific deployment can be interpreted as an instantaneous realization of an independent homogeneous Poisson point process (HPPP), represented by $\Phi_p$, with intensity $\lambda_p$ (expected number of aggregators per area unit). Since $ \Phi_{p}$ is a stationary process, the distribution of the points is invariant with respect to translation of the origin; therefore, the SIR analysis does not depend on the particular location of each aggregator. Thus, according to Slivnyak's theorem \cite{haenggi2012stochastic} and without loss of generality, we consider a “typical” aggregator located at the origin, which is subject to interference produced by the other non-intended transmitters in the network.\\
   Consider that at any instant, the MTDs across the entire network transmit information to their serving aggregators through the same set $\mathcal{N}$ of $N$ orthogonal channels. Each aggregator can accommodate only one MTD per channel, out of $K$ requesting service within its coverage area\textemdash $K$ is a Poisson distributed random variable with mean $m$, $K\sim {\operatorname {Poiss} }(m)$. Thus, the only contribution to the interference comes from the MTDs in the serving zones of other aggregators using the same channel (inter-cluster interference)\footnote{The probability that any MTD within any cluster generates interference does not depend on its position in the area respect the typical link\textemdash the channel occupation probability is $P_0=K/N$ when $N>K$ and $1$ otherwise. Thus, based on the independent thinning property, we can model the interference field observed from the typical link as an HPPP with density $P_0\lambda_p$. }. Notice that each MTD transmits whenever it has new information to send, and its corresponding aggregator has allocated resources for transmission\textemdash conforming the two scheduling schemes described in the following sections. Assuming that the MTDs have low mobility, complex association mechanisms between MTDs and aggregators are not needed; thus, we can model their locations as a Matérn cluster process (MCP)\footnote{This is a doubly Poisson cluster process that reflects in a better way the properties of the scenario treated in this work, compared to other point processes belonging to the same group such as the Thomas process \cite{haenggi2012stochastic}.}, where the aggregators form the parent point process \cite{guo2017massive}. 
  
  The MCP can be defined as
  \begin{equation}
  \renewcommand{\vec}[1]{\boldsymbol{#1}}
  \Phi\triangleq\bigcup\limits_{\mathbf{v}\in\Phi_p}\mathbf{v}+\mathbf{\mathcal{B}}^{\mathbf{v}},
  \end{equation}
  where $\mathbf{\mathcal{B}}^{\mathbf{v}}$ denotes the offspring point process; and each point $\mathbf{s}\in\mathbf{\mathcal{B}}^{\mathbf{v}}$ is independent and identical distributed around the cluster center $\mathbf{v}\in\Phi_p$ with distance distribution $f(r_d)=\frac{2r_d}{R_d^2}$, where $R_d$ is the radius of the clusters formed by the aggregators and its corresponding MTDs \cite{saha2017poisson}.
 Notice that the MCP definition implies that each MTD is associated with a single aggregator even though it might be the case that a particular MTD is located within several aggregators' coverage areas.  

We adopt a channel model that consists of the commonly used power-law path-loss as the large-scale propagation effect. Thus, the signal power decreases at a rate of $r^{-\alpha}$ with the propagation distance $r$, and $\alpha\geq2$ is the path loss exponent. Quasi-static fading is considered as the small-scale effect, which means that the channel is constant during a transmission block and changes independently from block to block. Additionally, Rayleigh multi-path fading environment is assumed, with intended and interfering channel power gains ($h$ and $g$, respectively) being exponentially distributed with unit mean. This allows us to examine the worst-case scenario without line of sight. 

All MTDs use full inversion power control. This is, each device controls its transmit power so that the average signal power received at the serving aggregator is equal to a predefined constant value $\rho$. This guarantees a uniform user experience while saving an important amount of energy \cite{lopez2018aggregation}. Due to the high density of MTDs and aggregators, we consider an interference-limited scenario (i.e., the co-channel interference limits the performance of all links, and the thermal noise at the receiver side can be neglected); consequently, the received SIR determines the network performance and the value of $\rho$ is irrelevant.
%
\section{The MD of the SIR}
The link success probability given a SIR threshold $\theta$, $p_{s}(\theta)=\mathbb{P}(\mathrm{SIR}> \theta)$, is a performance metric of interest in large-scale interference-limited networks. The computation of $p_s$ requires spatial averaging over the point process; thus, it does not reveal users with low success probability. In other words, $p_s$ allows designers to know the fraction of MTDs that succeed in transmitting, but it does not exhibit how concentrated the link success probabilities are; therefore, it is impossible to distribute the resources across the network properly. It is important to measure the fluctuation of the link reliability around $p_s$ to fully characterize the network's performance in terms of connectivity, end-to-end delay, and QoS. Thus, we center our attention on random variables of the form

\begin{equation}
 P_{s}(\theta)\triangleq\mathbb{P}(\mathrm{SIR} > \theta\vert\Phi) ,\label{1}
\end{equation}
where the conditional probability is taken over the fading and the channel access scheme and given the nodes' position for a particular realization of the network. Following this notation, the standard success probability would be $p_{s}(\theta)=\mathbb{E}[P_{s}(\theta)]$. The intention is then to find  the two-parameter complementary cumulative distribution function (CCDF) of $P_{s}(\theta)$, defined as \cite{haenggi2015meta}
\begin{equation}
 \bar{F}(\theta,x)\triangleq\mathbb{P}(P_{s}(\theta) > x),   
 \label{2}
\end{equation}
where $x\in[0,1]$ refers to the target reliability level. Due to the ergodicity of the point process, one can understand $ \bar{F}(\theta,x)$ as the fraction of links or users that achieve an $\mathrm{SIR}$ $\theta$ with probability at least $x$ \cite{tang2020meta}. Direct calculation of (\ref{2}) is usually infeasible because the probability density function (PDF) of the interference power is hard to find except for just a few path loss exponents and fading models. Consequently, some conjectures and approximations are usually made to ease the heavy mathematical work; but this frequently results in an inappropriate modeling of practical scenarios.

Different methodologies have been recently proposed in the literature to find closed-form expressions or approximate the MD. In \cite{8421026}, the entire MD is reconstructed only from its moments using Fourier-Jacobi expansions. Authors of \cite{haenggi2015meta} compare the results for the MD obtained from using the Gil-Pelaez theorem, the beta approximation, and Paley-Zygmund bound\footnote{ The Paley-Zygmund bound is useful to determine the fraction of links that attains at least a certain fraction of the average performance.}. Finally, an efficient calculation method based on binomial mixtures was presented in \cite{8374969}. This approach permits us to extract the MD from a linear transformation of the moments (simple matrix-vector multiplication). This linear mapping is an upper triangular matrix, symmetric with respect to the antidiagonal, and needs to be calculated only once for the desired level of accuracy. However, high order moments are needed, which is not always feasible to compute.

Most of these methods require the computation of complex integrals and are sensitive to the parameters selection. Differently, in the next section we provide a simple, yet efficient, method for computing the SIR MD under RRS. 

\section{Random Resource Scheduling (RRS)}

Under RRS, each aggregator randomly assigns the channels in $\mathcal{N}$ to the MTDs. Notice that this mechanism does not need channel state information (CSI). 
Since the MTDs use inversion power control, the $\mathrm{SIR}$ experienced by the typical user is $\mathrm{SIR}=\frac{h}{I}$, where $I=\sum_{i\in\Phi\backslash \{0\}} g_i r_{d_i}^{\alpha}y_i^{-\alpha}$ is the aggregated interference from MTDs in other clusters transmitting over the same channel, $\{y_i\}$ denotes the distance of the interfering MTDs respect the typical user, $h$ and $\{g_i\}$ are the fading power gains on the desired and interfering links, respectively, and $\{r_{d_i}\}$ is the distance between the MTDs and their serving aggregators. For an arbitrary but fixed realization of $\Phi$, the conditional success probability can be obtained  from (\ref{1})  as
 \begin{align}\label{4}
  P_{s}(\theta) & =  \mathbb{P}\left( \frac{h}{\sum\limits_{i\in\Phi\backslash \{0\}}g_i r_{d_i}^{\alpha}y_i^{-\alpha}} \geq\theta\middle|\Phi\right)\notag\\
  & =\mathbb{E}_{g_i}\left[ \mathbb{P}\left(h\geq\theta\sum_{i\in\Phi\backslash \{0\}}g_i r_{d_i}^{\alpha}y_i^{-\alpha}\middle| g_i,\Phi\right)\right]\notag\\ 
  &\stackrel{(a)} =
  \mathbb{E}_{g_i}\left[\exp\left(-\theta\sum_{i\in\Phi\backslash \{0\}}g_i r_{d_i}^{\alpha}y_i^{-\alpha}\right)\middle| \Phi\right]\notag\\&\stackrel{(b)}=\!\!\!\!\!\!
  \displaystyle\prod_{i\in\Phi\backslash \{0\}}\!\!\!\left[\frac{1}{1+\theta(\frac{r_{m_i}}{y_i})^\alpha}\right]\!\!\!\stackrel{(c)}=\!\!\lim_{\eta\rightarrow\infty}\displaystyle\prod_{i=1}^{\eta}\left[1+\theta\left(\frac{r_{d_i}}{y_i}\right)^\alpha\right]^{-1}\notag\\
&\stackrel{(d)}\leq\lim_{\eta\rightarrow\infty}\left[\frac{\sum_{i=1}^{\eta}\left[1+\theta\left(\frac{r_{d_i}}{y_i}\right)^\alpha\right]}{\eta}\right]^{-\eta}\nonumber\\
&\stackrel{(e)}=\lim_{\eta\rightarrow\infty}\!\left[1+\frac{\theta}{\eta}\sum_{i=1}^{\eta}\left(\frac{r_{d_i}}{y_i}\right)^\alpha\right]^{-\eta}\!\!\!\!\!=\lim_{\eta\rightarrow\infty}\left[1+\frac{\theta}{\eta}\beta\right]^{-\eta}\notag\\&\stackrel{(f)}\approx e^{-\beta\theta},
\end{align}
 where $(a)$ comes from applying the CCDF of the unit mean exponential distribution of $h$; $(b)$ follows from using the Laplace transform of the sum of independent and identical distributed exponential random variables $\{g_i\}$; in $(c)$, we consider the number of interfering nodes as $\eta$, and then use the relation between the geometric and the arithmetic mean to obtain $(d)$. Note that interchanging geometric and arithmetic means was shown to conduce to accurate performance expressions in \cite{lopez2018rate} for sufficiently dense networks without aggregation. Finally, $(e)$ comes from simple algebraic transformations; and $(f)$ results from making $\beta=\sum_{i=1}^{\eta}\left(\frac{r_{d_i}}{y_i}\right)^\alpha$.
 
\begin{remark} 
 Obtaining the PDF of $P_{s}(\theta)$ directly from ($b$) seems infeasible for massive network deployments; thus, in this case, we adopt approximation (\ref{4}) to attain the final SIR MD analytical expression. 
\end{remark}
The Laplace transform of $\beta$ (See Appendix A) has the form of a stretched exponential or Kohlrausch function, i.e, $e^{-ts^{2/\alpha}}$ for a certain constant $t$. In  \cite[Table I]{ammar2018closed}, the authors provide the PDF of such random variables for different values of $\alpha$. Herein, we adopt these results and set $\alpha$ to 4 for simplicity. Then, the PDF of $\beta$ is given by \cite[Table I]{ammar2018closed}
\begin{align}\label{PDF}
f_{\beta}(\omega)=\frac{t e^{\frac{-t^2}{4\omega}}}{2\sqrt{\pi}\omega^{\frac{3}{2}}}.
\end{align}
 From (\ref{2})-(\ref{PDF}), we attain the SIR MD under RRS as
\begin{align}\label{6}
\bar{F}(\theta,x)\!&=\!\mathbb{P}\left(e^{-\beta\theta}\!>\!x\right)\!=\!\mathbb{P}\left(\beta\!<\!\frac{-\ln{x}}{\theta}\right)\!=\!\int_{0}^{\frac{-\ln{x}}{\theta}} f_{\beta}(\omega)d\omega\notag\\
& =\int_{0}^{\frac{-\ln{x}}{\theta}}\frac{t e^{\frac{-t^2}{4\omega}}}{2\sqrt{\pi}\omega^{\frac{3}{2}}}d\omega\nonumber\stackrel{(a)}=b\int_{\frac{-\theta}{\ln{x}}}^{\infty}u^{\frac{1}{2}-1}e^{-au}du\notag\\
&=b a^{-\frac{1}{2}}\Gamma\left(\frac{1}{2},-\frac{\theta}{\ln{x}}\right),
\end{align}
where $(a)$ comes from substituting $\tfrac{1}{\omega}\!=\!u$,  $a\!=\!\tfrac{t^2}{4}$, $b\!=\!\tfrac{t}{2\sqrt{\pi}}$ and $t=\tfrac{1}{2}P_0\lambda_p\pi R_d^2\Gamma\left(1-\frac{2}{\alpha}\right)$. Note that $P_0\lambda_p$ is the density of the served MTDs and $P_0$ is the average channel occupation probability, which is independent of the scheduling scheme \cite{guo2017massive}:
\begin{align}
P_0&=\mathbb{E}_{K}\left[\frac{\min(K,N)}{N}\right]=\sum_{k=0}^{N}\frac{k}{N}\frac{m^k e^{-m}}{k!}\!+\!\!\sum_{k=N+1}^{\infty}\!\frac{m^k e^{-m}}{k!}\nonumber\\
&=1-\frac{\Gamma[1+N,m]}{N!}+\frac{m\Gamma[1+N,m]-e^{-m}m^{N+1}}{(N-1)!N^2}.
\label{7}
\end{align}
From the previous equation it becomes evident that as $N$ increases, the occupation probability for a certain channel decreases, which improves the performance.
\begin{remark}
The same procedure to compute the SIR MD in (\ref{6}) applies for any integer value of path loss exponent $\alpha>2$.
\end{remark}
\section{Channel-aware Resource Scheduling (CRS)}
Contrary to the RRS scheme, the aggregators that implement CRS allocate the available channel resources to the MTDs with better $\mathrm{SIR}$ (equivalently, better channel gains). Herein, each aggregator is assumed with perfect CSI of its associated MTDs. From (\ref{1}), we have that the conditional success probability corresponding to the $\nu$-th link with $h(\nu)$ channel
power gain is
\begin{align}
 P_s(\theta)\triangleq\mathbb{P}(\mathrm{SIR} > \theta\vert\Phi)=1-\mathbb{P}(h(\nu)<\theta I\vert\Phi),\label{8}
\end{align}
where $\nu=1,...,N$ corresponds to the selected MTDs ordered according their $\mathrm{SIR}$ as $h(1)\!>\!\ldots\!>\!h(\nu)\!>\!\ldots\!>\!h(N)$. Then, the cumulative distribution function (CDF) of $h(\nu)$ using order statistics results is $F_{h(\nu)}(v)\!=\!\sum_{l=q}^{K}\!\binom{K}{l}[F_h(v)]^l[1\!-\!F_h(v)]^{K\!-\!l}$, where $q=K-\nu+1$ and $F_h(v)$ is the CDF of an exponential random variable. The methodology to find the SIR MD is the same as in Section \textrm{III}:
\begin{align}\label{9}
 &\bar{F}(\theta,x)\nonumber\\
 &\ \stackrel{(a)}=\!\mathbb{P}\bigg(1\!-\!\sum_{l=q}^{K} \binom{K}{l}\mathbb{E}_{I}\!\left[ [1-e^{-I\theta}]^l
 e^{-I\theta (K-l)}\middle|\Phi\right]>x\bigg)\notag\\
 &\ \stackrel{(b)}=\!\mathbb{P}\bigg(1\!-\!\!\sum_{l=q}^{K}\!\sum_{r=0}^{l}\! \binom{K}{l}\!\binom{l}{r}\!(-1)^r\mathbb{E}_{I}\!\left[ \!e^{-I\theta(K-l+r)}\!\middle|\Phi\!\right]\!>\!x\!\bigg)\notag\\
 &\ \stackrel{(c)}=\!\mathbb{P}\bigg(\!1\!-\!\!\sum_{l=q}^{K}\!\sum_{r=0}^{l}\! \binom{K}{l}\!\binom{l}{r}\!(-1)^r\!\!\prod_{i\in\Phi_I}\!\!\frac{1}{1\!+\!\theta_{l,r}(\frac{r_{d_i}}{y_i})^\alpha}\!>\!x\!\bigg)\!,
\end{align}\\
 where $(a)$ comes from combining (\ref{8}) and (\ref{2}), and from using the CDF of $h(\nu)$; $(b)$ is obtained by using the binomial expansion $(1\!+\!z)^l\!=\!\sum_{r=0}^{l} \binom{l}{r}z^r$; and ($c$) follows from considering the expectation in $(b)$ as the Laplace transform of the interference conditioned on the point process, and by making $\theta_{l,r}\!=\!\theta(K\!-\!l\!+\!r)$. Herein, we analyze the worst-case performance, which is when $\nu\!=\!N$. This is, we consider the link with the smallest channel coefficient among all the links with access granted by the aggregator. Thus, by setting $q\!=\!K\!-\!N\!+\!1$ in (\ref{9}), we avoid averaging over all possibilities of $\nu$, reducing complexity while keeping the per-link performance guarantee concept. Notice that considering all choices of $\nu$ would lead to an average result, which contradicts the MD concept. Moreover, (\ref{9}) allows a semi-analytical computation of the MD that depends only on the interfering nodes' position inside the clusters and with respect to the typical link. Therefore, it is unnecessary to model either the channel fading or the scheduling process, which reduces the computation time significantly. %
\section{Numerical Results}
This section presents numerical results to analyze the per-link performance in an mMTC setup with data aggregation under RRS and CRS. The aggregators are deployed in a disk of radius 3 km with density $\lambda_p=3\times 10^{-6}$ aggregators/$\mathrm{m}^2$, which guarantees  100 aggregators deployed on average in the area while eliminating the impact of the border effect in the simulation. It was selected $N=20$, $m=60$, and $R_d=40$ m to produce visualization errors in the order of $10^{-2}$. Monte Carlo based results are obtained with $10^5$ samples and are included in the figures with markers to validate our analytical (for RRS) and semi-analytical (for CRS) expressions, represented with lines.
  
\begin{figure}[h]
    \centering
    \includegraphics[width=\columnwidth]{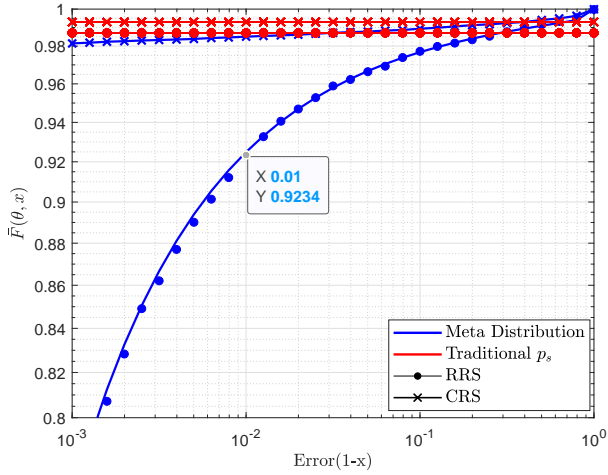}
    \vspace{-4mm}
    \caption{Meta distribution and the traditional success probability ($p_s$) for $\alpha=4$ and $\theta=0$ dB. The marked point can be interpreted as approximately $92\%$ of the users can communicate with 0.01 probability of error.}
    \vspace{-4mm}
    \label{fig2}
\end{figure}

 Fig.~\ref{fig2} shows the efficacy of the SIR MD for describing the system per-link performance. One can realize that the traditional $p_s$ does not guarantee QoS for any node in the network, which is even more remarkable when RRS is enabled in the aggregation phase. This is because the channels are assigned to links that communicate with high error probability. In contrast, having the MD in hand allows effective distribution of those resources as the exact fraction of links communicating with a target reliability is known in advance. Moreover, aggregators implementing CRS admit a higher percentage of links achieving a target reliability in the resources-constrained communication system.
\begin{figure}[t!]
    \centering
    \includegraphics[width=\columnwidth]{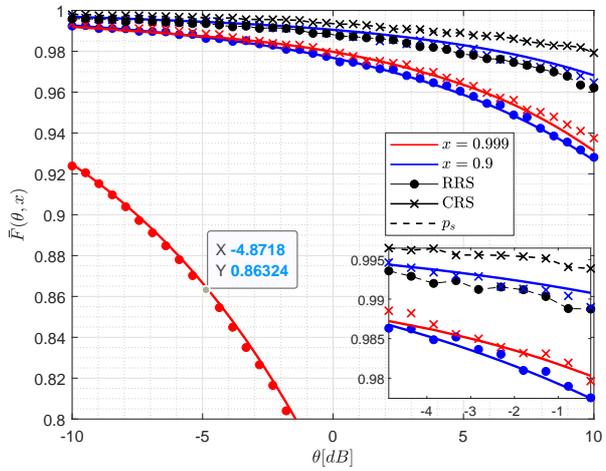}
    \vspace{-4mm}
    \caption{Meta distribution and traditional success probability ($p_s$) as a function of $\theta$ for different values of reliability ($x$).}
    \vspace{-4mm}
    \label{fig3}
\end{figure}

Fig.~\ref{fig3} permits a more rigorous analysis of the fraction of links that achieve certain reliability given the $\mathrm{SIR}$ threshold $\theta$. For example, the marked point shows that the transmission rate should be set no greater than $\log_2(1+10^{-5/10})=0.396$ bits per channel use (bpcu) to guarantee nearly $86\%$ of the links achieving at least $99.9\%$ success probability. One may notice that nearly the same fraction of devices can transmit with the same rate under both scheduling schemes, but with reliability improved from 0.9 when using RRS to 0.999 under CRS. If simplicity is desired in the network, RRS is the solution, but only a small percentage of devices achieve high reliability. However, if a larger number of devices need to communicate, CRS seems to be the best option to provide them the required reliability.
\begin{figure}[t!]
    \centering
    \includegraphics[width=\columnwidth]{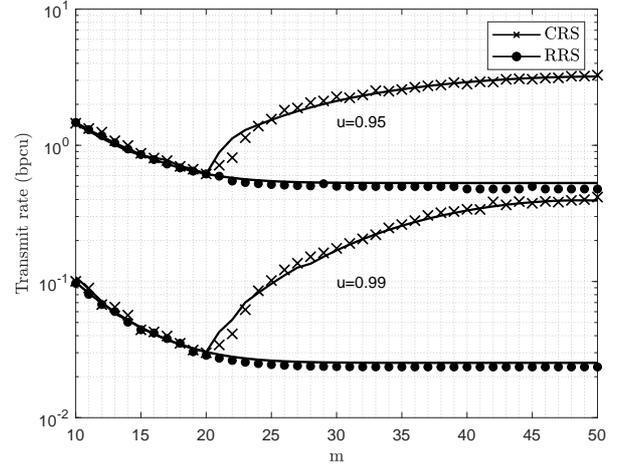}
    \vspace{-4mm}
    \caption{Transmit rate versus average number of MTDs per aggregator for $\bar{F}(\theta,x)=u=0.99,0.95$ and $x=99\%$ of reliability.}
    \vspace{-4mm}
    \label{fig4}
\end{figure}

Fig.~\ref{fig4} visualizes the trade-off between the transmit rate and $m$. In general, the transmit rate has to decrease when the number of MTDs requesting transmission approaches the available resources. However, when this value exceeds $N$, the probability of selecting a better channel under CRS increases, and the transmit rate can improve. In contrast, under RRS the rate can not exceed a constant value. Notice that the percentage of devices communicating with the target reliability must remain the same in both cases.

\begin{figure}[t!]
    \centering
    \includegraphics[width=\columnwidth]{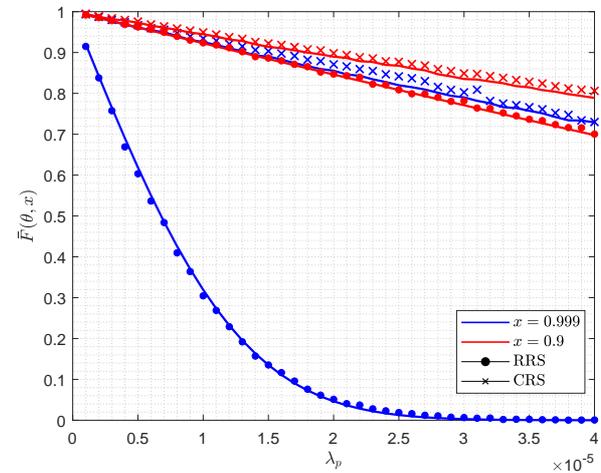}
    \vspace{-4mm}
    \caption{Meta distribution as a function of $\lambda_p$ for $\alpha = 4$, $\theta=0$dB and $x=0.999,0.9$.}
    \label{fig5}
    \vspace{-4mm}
\end{figure}

Fig.~\ref{fig5} shows an important trade-off that network designers need to consider. This is, the aggregators' density can not increase loosely; otherwise, only a small fraction of MTDs would communicate with the target reliability. On the other hand, it is clear that as $\lambda_p$ approaches $0$, the highest percentage of MTDs achieving the target reliability, but this is not the best choice because we would not benefit from aggregation. Notice that, in general, the density of served MTDs increases with $\lambda_p$. However, under RRS, almost no transmissions satisfy 99\% of reliability. Moreover, based on this figure results, one can determine the density of MTDs that allows reaching an SIR  $\theta$ with reliability $x$, which is given by $\lambda_p P_0 \bar{F}(\theta,x)$.

\section{Conclusions}
In this work, we analyzed the meta distribution of the $\mathrm{SIR}$ to provide a more fine-grained description of the per-link performance under RRS and CRS scheduling schemes compared to the analyses derived from the traditional success probability.
Our results showed that RRS performs extremely poor in terms of per-link reliability compared to CRS, when a limited amount of resources can be scheduled among the MTDs. This is, RRS guarantees the same per-link reliability performance as CRS but for a smaller number of communication links. This difference is more significant when targeting stringent communication errors. In future works, we might explore analytical approaches to characterize the meta distribution under CRS and implement an algorithm that ensures efficient power and rate control in these challenging scenarios. 
\section*{Acknowledgment}
This work is supported by Academy of Finland (Grants n.307492 Academy Professor, n.318927 6Genesis Flagship,  n.326301 FIREMAN and  no319008 EE-IoT). 

\appendices
\section{Laplace transform of $\beta$}
We proceed as follows
\begin{align}\label{13}
\mathcal{L}_{\beta}(s)&\!=\!\mathbb{E}_{\Phi,r_{m}}\bigg[\exp{\Big(-s\sum_{i\in\Phi\backslash \{0\}}\big(\tfrac{r_{d_i}}{y_i}\big)^\alpha\Big)} \bigg]\\
&=\mathbb{E}_{\Phi}\prod_{i\in\Phi\backslash \{0\}}\mathbb{E}_{r_{d}}\Big[\exp{\Big(-s\big(\tfrac{r_{d_i}}{y_i}\big)^\alpha\Big)}\Big]\nonumber\\
&\stackrel{(a)}=\!\exp\bigg(\!\!-\!2\pi\lambda\!\int_{0}^{\infty}\!\!\!\!\bigg(1\!-\!\mathbb{E}_{r_{d}}\Big[\exp\Big(\!-s\big(\tfrac{r_{d}}{y}\big)\!^\alpha\Big)\Big]\bigg)ydy\bigg)\nonumber\\
&\stackrel{(b)}=\!\exp\bigg(\tfrac{-2\pi\lambda}{\alpha}\mathbb{E}_{r_{d}}\int_{0}^{\infty}\left[1-\exp\left(\tfrac{-sr_{d}^\alpha}{w}\right)\right]w^{\frac{2}{\alpha}-1}dw\bigg)\nonumber\\
&\stackrel{(c)}=\exp\bigg(-\pi P_0\lambda_p\mathbb{E}_{r_{d}}\Big[\mathbb{E}_{w}\Big[\big(\tfrac{w}{(sr_{d})^\alpha}\big)^{-\frac{2}{\alpha}}\Big)\Big]\bigg)\nonumber\\
&=\exp\Big(-\pi P_0\lambda_p\mathbb{E}_{r_{d}}[r_{d}^2]\ \mathbb{E}_{w}\big[w^{\frac{2}{\alpha}}\big]\Big)\nonumber\\
&=\exp\Big(\!\!\!-\!\pi P_0\lambda_p\tfrac{R_{d}^2}{2}\Gamma\!\!\big(1-\tfrac{2}{\alpha}\big)s^{\tfrac{2}{\alpha}}\!\Big)\!=\exp\big(\!\!-\!ts^{\tfrac{2}{\alpha}}\big),
\end{align}
where $(a)$ comes from applying the probability generating functional (PGFL) defined in \cite{haenggi2009interference},\cite{yu2019stochastic}; $(b)$ follows from the substitution $y^{-\alpha}=\frac{1}{w}$; $(c)$ is obtained using the following definition of moment of a non-negative continuous real random variable $\mathbb{E}[z^n]=\int nz^{n-1}(1-F(z))dz$, and by noticing that $\mathbb{E}_{w}\left(w^{\frac{2}{\alpha}}\right)$ is the expected value of an exponential random variable with unitary mean. \hfill 	\qedsymbol 

\bibliographystyle{IEEEtran}
\bibliography{References}

\end{document}